\documentstyle[12pt]{article}

\newcounter{multieqs}

\newcommand{\bq}{\begin{equation}}
\newcommand{\fq}{\end{equation}}
\newcommand{\bqr}{\begin{eqnarray}}
\newcommand{\fqr}{\end{eqnarray}}
\newcommand{\non}{\nonumber \\}
\newcommand{\noi}{\noindent}

\newcommand{\rf}[1]{(\ref{#1})}

\def\alp{\alpha}
\def\bet{\beta}
\def\gam{\gamma}
\def\del{\delta}
\def\eps{\epsilon}
\def\veps{\varepsilon}

\def\th{\theta}

\def\kap{\kappa}
\def\lam{\lambda}

\def\sig{\sigma}

\def\Gam{\Gamma}

\def\cF{{\cal F}}

\def\cN{{\cal N}}

\def\cU{{\cal U}}

\def\pa{\partial}

\def\pr{^{\prime}}
\def\hlf{\frac{1}{2}}
 
\def\rar{\rightarrow}

\begin{document}

\thispagestyle{empty}
\setcounter{page}{0}

%%%%%%%%%%%%%%%%%%%%%%%%%%

\begin{flushright}
\begin{tabular}{l} 

ANL-HEP-99-78 \\
YITP-99-48  \\
hep-th/9909087  \\ 
\end{tabular}
\end{flushright}

\vspace{9mm}
\begin{center}

{\Large \bf Quantization and Scattering in the IIB $SL(2,Z)$ 
Covariant Superstring}\\

\vspace{12mm}

{Gordon Chalmers${}^{\dagger}${}\footnote{E-mail address: 
chalmers@pcl9.hep.anl.gov} and Koenraad Schalm${}^*$
}\footnote{E-mail address: konrad@insti.physics.sunysb.edu. \\
\hskip .4in Address after September 15: NIKHEF, P.O Box 41882, 1009 DB Amsterdam.}
\\[10mm]
${}^{\dagger}${\em Argonne National Laboratory \\
High Energy Physics Division \\
9700 South Cass Avenue \\
Argonne, IL  60439-4815 } \\[5mm]
${}^*${\em C.N. Yang Institute for Theoretical Physics  \\
State University of New York at Stony Brook  \\
Stony Brook, NY 11794-3840 }

\vspace{5mm}

{\bf Abstract}
\end{center}
We rewrite the $SL(2,Z)$ covariant worldsheet action for the IIB string
proposed by Townsend in a Polyakov form.  In a flat background the formalism 
yields separate $(p,q)$ sectors.  In each one the action is that of 
the IIB string action with the string slope parameter $\alp\pr$ 
replaced by its $SL(2,Z)$ analogue $\alp_{pq}\pr$.  $SL(2,Z)$ invariant 
graviton scattering amplitudes are obtained from those of 
the fundamental $(1,0)$ string by summing over the different sectors.  
The tree-level four-graviton amplitude in this formalism differs 
from a previously conjectured non-perturbative form; both yield the 
same expansion in order $\alpha'^3$.  
\vfill

\newpage
\setcounter{footnote}{0}
\section{Introduction}  

Exploring the non-perturbative structure of the IIB superstring 
is relevant for an understanding of its dynamics and for a potential  
formulation of the conjectured M-theory.  Progress in 
the pinning down of the derivative expansion of the IIB effective action has  
been made in the recent past, especially with regards to the conjectured 
self-duality of the IIB theory.  Relevant to this 
program is an attempt to reorganize the perturbative expansion in a manifestly 
$SL(2,Z)$ invariant fashion and to find further contributions or structure 
in the derivative expansion. 

The conjectured $SL(2,Z)$ S-duality of IIB string theory 
maps the fundamental string into a solitonic D-string. This led 
Townsend to directly incorporate the self-duality at the string 
level in an $SL(2,Z)$ covariant worldsheet 
action for the bosonic fields of the superstring \cite{Townsend:1997}.
A subsequent supersymmetric version was constructed by him together
with Cederwall \cite{Cederwall:1997}.  These actions are special in
that the tension is 
generated dynamically by a two-dimensional worldsheet gauge 
field that has no local degrees of freedom.  Classically these fields
may be 
integrated out and the solution to the field equations contains an 
integration constant which acts as the tension. It is the presence of
this extra worldsheet gauge field that allows one to covariantize the
action with respect to the conjectured 
$SL(2,Z)$ S-duality. One may combine it with the Born-Infeld gauge
field of the D-string into an $SL(2,Z)$ doublet and a 
Hamiltonian analysis then shows that the tension  
of this self-dual string is promoted to its $SL(2,Z)$ analogue
\cite{Townsend:1997, Schwarz:1995, Schwarz:1996}  
\bq
T_{pq} = \frac{1}{2 \pi \alp\pr_{pq}} = \frac{1}{2\pi\alp\pr}
\sqrt{ q^2 e^{-2\phi}+(p+q\chi)^2} \ .
\label{tpq} 
\fq
Here $\phi$ and $\chi$ denote the vacuum expectation values of the
dilaton and axion respectively (the string coupling is $g_s=e^\phi$).   

In a separate development Russo has made an ansatz for the 
non-perturbative four-graviton amplitude of the ten-dimensional 
IIB superstring theory \cite{Russo:1997, Russo:1998}.  An origin of 
this ansatz lays in an $SL(2,Z)$ invariant set of corrections 
to the IIB low energy effective action at order $R^4$ in derivatives  
\cite{Russo:1997mk, Green:1997one, Green:1997two, Green:1997three, 
GreenSethi,kehagias,Kehagias:1997}.  As the
non-perturbative action of $S$-duality mixes the weak and strong 
coupling regimes of the string, these results generate 
full non-perturbative answers for the low energy process of graviton 
scattering.  An interesting feature of Russo's ansatz is that it may 
be obtained from the standard four-graviton amplitude by replacing the 
tension with its $SL(2,Z)$ covariant analogue $T_{pq}$ of equation 
\rf{tpq} provided one incorporates the different $(p,q)$ sectors in an 
appropriate way.  This has led to the speculation that this amplitude 
may be computable from a $SL(2,Z)$ covariant worldsheet action.   

In this letter we discuss quantization of 
this worldsheet action.  By the appropriate 
introduction of auxiliary fields we may 
linearize the Nambu-Goto form of the $SL(2,Z)$ covariant action 
to a Polyakov form, from which we know how to compute explicit
amplitudes for the fundamental string.  We will see that in the $SL(2,Z)$ 
covariant Polyakov formulation the field strengths  
of the required $SL(2,Z)$ worldsheet gauge fields appear linearly as
they do in the manifestly four-dimensional formulation in \cite{Berkovits}.  
New is the presence of a Lagrange multiplier field which forces the 
whole Lagrangian to vanish.  The Lagrange
multiplier times linear field strength  
term is familiar from Lagrangian formulations
of $T$-duality \cite{Buscher:1987, 
Buscher:1988, Rocek:1992}.  In particular gauge invariance forces the
constant zero mode of the Lagrange multiplier field to be quantized 
\cite{Rocek:1992, Witten:1996,Schmidhuber}.  Furthermore the constraint 
implied by the linear appearance of the gauge field requires the 
non-zero modes to vanish.  As this Lagrange multiplier appears as 
an overall term in front of the whole Lagrangian its constant part 
in effect becomes the $SL(2,Z)$ covariant tension $T_{pq}$. The 
remaining action is that of the free string.  We thus establish that 
amplitudes in the $(p,q)$ sector are given by those of the 
fundamental string after substitution with $T_{pq}$, as is 
required by S-duality. 

For the $SL(2,Z)$ covariant superstring the analysis is identical and 
yields the Green-Schwarz string with tension $T_{pq}$, but as usual
kappa-symmetry prevents covariant quantization.  The believed 
equivalence between the GS string and the RNS formalism allows us to 
posit the use of the latter instead.

The tension arises dynamically in the $SL(2,Z)$ covariant formulation and
each of the $(p,q)$ strings is a groundstate of the worldsheet theory. The
worldsheet path integral then requires the sum over each of these
different sectors. The use of the naive perturbative string vertex
operators in each sector yields $SL(2,Z)$ covariant amplitudes as
the sum over the individual ones from each $(p,q)$ string. The amplitude
for four-graviton scattering derived in this way reproduces the complete
$R^4$ term in the IIB low energy effective action as does the
conjecture of Russo. In summing over the amplitudes of the
$(p,q)$ strings one counts the massless graviton tree exchange an 
infinite number of times, but this has no effect on the $R^4$ higher 
derivative term in the effective action; that the infinite
summation over the $(p,q)$ strings gives rise to the $R^4$
term was noticed in \cite{kehagias}. The conjectured form of Russo on the
other hand resembles the product of the amplitudes over the different
sectors and contains just a single massless pole. A possible
modification of the vertex operators with factors depending on the
worldsheet gauge fields is unlikely to change the sum over the $(p,q)$
sectors into a product. Aside from the number of massless poles,
which amplitude yields the correct non-perturbative information for the
IIB string can only be determined at the higher-derivative level.

The amplitude derived here is a tree-level result in the worldsheet genus
expansion of the $SL(2,Z)$ covariant string. Russo has conjectured that
the product form should receive $SL(2,Z)$ covariant corrections as well.
In the complete form of the non-perturbative IIB graviton scattering
amplitude one would expect cuts required by unitarity as well as imaginary
parts arising from poles off the real axis due to the unstable states of
the $(p,q)$ strings. The problem for both $SL(2,Z)$ covariant amplitudes
is identifying the correct small parameter which governs the perturbative
expansion of the $SL(2,Z)$ covariant string; for each $(p,q)$ string the
natural expansion parameter is the $(p,q)$ dilaton, which measures the
topology of the Riemann surface.  
There does exist a proposal by Berkovits for a
manifestly S-dual worldsheet action for the IIB string on a Calabi-Yau
three-fold which contains a coupling to the worldsheet curvature
\cite{Berkovits}. 

The organization of this work is as follows. First we derive a Polyakov
form of the manifestly $SL(2,Z)$ covariant string. In section 3 we solve
the constraints arising from the field equations and show that the
resulting action is that of separate $(p,q)$ strings. The IIB superstring
is presented in section 4. In section 5 we generate amplitudes for the
$SL(2,Z)$ covariant superstring and analyze the results.  

\section{Polyakov action for the $SL(2,Z)$ covariant IIB string}
\setcounter{equation}{0}

We first briefly review the origins of the action that lends itself 
to an $SL(2,Z)$ covariant formulation \cite{Townsend:1997,Bergshoeff:1992}.  
This formulation is in terms of an 
action where the tension is generated as a dynamical variable 
\cite{Townsend:1997, Bergshoeff:1992}, 
\bq
S_T=\int d^2\sig~ \frac{\lam^2}{2} \left[{\det(G_{\alp\bet}) 
 \over (2\pi\alpha')^2} + (\eps^{\alp \bet} F_{\alp\bet})^2\right] \ .
\label{act1} 
\fq 
Here $\epsilon^{\alpha\beta}$ is the Levi-Civita symbol in two 
dimensions.  The field $\lam (\sig)$ transforms under worldsheet coordinate 
transformations as $\lam  \rar \lam\pr = \lam \vert \pa \sig\pr/ 
\pa \sig\vert^{-1/2}$;
$G_{\alp\bet}$ is the pullback of the space-time metric $G_{\alp\bet}
= G_{\mu\nu}\pa_{\alp}X^{\mu}\pa_{\bet}X^{\nu}$ and $F_{\alp\bet}=
\pa_{[ \alp}A_{\bet ]}$ a worldsheet field strength. The equation of
motion for $A_{\gam}$ reads $\pa_{\gam} (\lam^{2} \eps^{\alp \bet}
F_{\alp\bet})=0$. After substituting the general solution $\eps^{\alp \bet}
F_{\alp\bet} = T/\lam^2$ for constant tension $T$ back into the action one
recovers the Schild action. This reduces to the
Nambu-Goto form after one integrates out the remaining auxiliary field
$\lam$.  The action \rf{act1} is in
addition invariant under the rigid spacetime scale variations
\bqr 
X \rar X\pr =e^t X, & \lam \rar \lam\pr = e^{-2t}\lam, & A_{\alp} \rar
A_{\alp}\pr = e^{2t}A_{\alp} \ .
\fqr
This invariance was a motivation behind its construction \cite{Bergshoeff:1992}.

The appearance of the extra worldsheet gauge field $A_{\gam}$ makes the
above action a starting point for a formulation of the $SL(2,Z)$
covariant IIB string. A comparison 
of the regular string action,  
\bq
S_F = {1\over 2\pi\alpha'} \int d^2\sig \sqrt{|G_{\alp\bet}|} + 
 {1\over 2\pi\alpha'}\int d^2\sig~ \eps^{\alp\bet}B_{\alp\bet}^{NS} \ ,
\fq
with the Born-Infeld action of the D-string,
\bq
S_D = T_{D1}\int d^2\sig \sqrt{|G_{\alp\bet} + 2\pi\alpha'
F_{\alp\bet}+B_{\alp\bet}|} + T_{D1}\int d^2\sig~ 
\eps^{\alp\bet}B_{\alp\bet}^{RR} \ ,
\fq
reveals that the absence of a gauge field on the worldvolume of the
fundamental string is a fundamental difference between the two. If
one disregards considerations of the D-string action as a solely
``effective one'' and implements S-duality at this level
there should be a way to also introduce a worldsheet gauge field for the
fundamental string, though without changing its dynamics. This is what the 
action~\rf{act1} accomplishes. The similarity between the D-string 
action and the action~\rf{act1} for the fundamental string
with dynamical tension can be made even more suggestive by realizing
\cite{Bergshoeff:1992} that the action~\rf{act1} is equivalent to the
Born-Infeld form 
\bq
S_T = \int d^2\sig ~\frac{\lam^2}{2(2\pi\alp\pr)^2} \det(G_{\alp\bet} +
2\pi\alpha' {\tilde F}_{\alp\bet}) \ .
\fq
The tilde over the field-strength is added to distinguish this worldsheet 
gauge field which generates the tension from the Born-Infeld gauge field 
of the D-string. 

To construct the $SL(2,Z)$ covariant action~\cite{Cederwall:1997} one thus
introduces two worldsheet gauge fields $A_{\alp}^{(i)}$ and combines them
into a doublet. Moreover, the field strengths are now both extended
with the doublet of pullbacks 
of the NS-NS and R-R antisymmetric tensor fields 
\bq
F^{(i)}_{\alp\bet} = \pa_{[\alp}A^{(i)}_{\bet]} + 
 \frac{B^{(i)}_{\alp\bet}}{2\pi\alp\pr} \ ,
\fq
to the well-known $F+B$ combination.

To complete the construction one also needs the background scalars of
the theory: the dilaton and the
axion. These belong to the coset $SL(2,R)/SO(2) \simeq SU(1,1)/U(1)$ 
and are denoted by the doublet $\cU_i$. They combine with the 
$F^{(i)}_{\alp\bet}$ into an $SU(1,1)$ invariant field strength  
\bq
\cF_{\alp\bet} = \cU_i F^{(i)}_{\alp\bet} \ .
\fq 
To relate this to a more familiar description of the background
scalars one may make the $U(1)$ gauge choice 
$\cU_1=e^{\phi/2}$ and $\cU_2 = e^{\phi/2}\chi+ie^{-\phi/2}$.   

The $SL(2,Z)$ covariant action is then given by, 
\bq 
S_{SL(2,Z)}= \int d^2\sig~ \frac{\lam^2}{2}\left[\frac{\det(G)}{(2\pi\alp\pr)^2}  
 + \Phi\bar{\Phi}\right].
\label{s1}
\fq 
Here $\mbox{det}(G)$ is the determinant of the pullback of the Einstein
frame metric; it is this metric which is $SL(2,Z)$
invariant. The field $\Phi$ is the Hodge dual 
of the $SU(1,1)$ invariant field strength: $\Phi= \eps^{\alp \bet} 
\cF_{\alp\bet}$.  

To be able to quantize the theory and compute scattering elements 
we would rather work with a Polyakov form of the action. 
In order to rewrite the above action in such a manner, we shall make 
use of the following exponential integral,  
\bq
\int_{-\infty}^{\infty}dx~ e^{-x^2-\frac{a^2}{x^2}} = \sqrt{\pi} e^{-2a} 
 \ ,
\label{bgmac}
\fq
to change the unit power determinant $\mbox{det}(G)$ to the square
root $\sqrt{\mbox{det}(G)}$ and obtain a Nambu-Goto form of
the string action. The above integral illustrates
the formal equivalence of the Schild action with the Nambu-Goto one. 
Eq. \rf{bgmac} is the reason for using a square for the worldsheet scalar 
field.  An analogous functional form of \rf{bgmac} is then also exact in the path 
integral. 

The first step is to obtain a term in the action that behaves as $1/\lam^2$.  
This is straightforward as the form of the gauge field couplings in 
\rf{s1} may be reproduced by the introduction of an auxiliary source 
field $k$ coupling to $\Phi$, whose kinetic term will then exhibit the
desired behaviour. 
\bq
S_{SL(2,Z)}= \int d^2\sig~ {\lam^2 \over 2(2\pi\alpha\pr)^2}\det(G) -
\frac{1}{2\lam^2}k\bar{k} +
\frac{1}{2}\left(\bar{k}\Phi+\bar{\Phi}k\right) \ . 
\label{chesburg}
\fq 
Integrating out $k$ gives back the previous $SL(2,Z)$ covariant action.  

We next integrate out the scalar field $\lam$ using the 
identity in eq.~\rf{bgmac}. This yields 
\bqr
S_{SL(2,Z)} &=& \int d^2\sig~ - \frac{1}{4\pi\alp\pr} 
  \sqrt{-\det(G)}\sqrt{k\bar{k}} +
\frac{1}{2}\left(\bar{k}\Phi+\bar{\Phi}k\right) \ .
\fqr
Redefining $k= \rho e^{i\th}$, we obtain the Nambu-Goto
form of the IIB $SL(2,Z)$ covariant action 
\bqr
S_{SL(2,Z)}&=& \int d^2\sig~ -
  \frac{\rho}{4\pi\alp\pr} \sqrt{-\det(G)}+
\frac{1}{2}\left(\rho e^{-i\th}\Phi+\rho\bar{\Phi}e^{i\th} \right) \ .
\label{finform}
\fqr
Rescaling the field $\rho \rar (2\pi\alp\pr)\rho$ does not eliminate 
its effect in the action because the $\Phi$ dependent terms will pick 
up this factor.  (We ignore the Jacobian effects due to the field 
redefinitions since we are examining classical actions.) 

It is now straightforward to introduce an auxiliary worldsheet metric
$g_{\alp\bet}$ and obtain the Polyakov form of the action. The 
scalar field $\rho$ enters multiplicatively in \rf{finform} and
therefore does not alter the form of the worldsheet metric field
equations.  The gauge field
terms involving $\Phi$ are also worldsheet coordinate invariant 
provided the epsilon used to contract the field strength is the density 
and not the tensor
\bqr
\Phi = \eps_{\mbox{\footnotesize
Levi-Civita}}^{\alp\bet}\cF_{\alp\bet}= \sqrt{g}
\veps^{\alp\bet}\cF_{\alp\bet} \ .
\fqr  
The Polyakov form of the action is thus
\bqr
S^{Pol}_{SL(2,Z)}= \int d^2\sig~ \frac{\rho}{4\pi\alp\pr} \sqrt{g}  
 g^{\alp\bet} G(X)_{\mu\nu}\pa_{\alp}X^{\mu}\pa_{\bet}X^{\nu}
 - \hlf\int d^2\sig
 \left[ \rho e^{-i\th}\Phi + \rho\bar{\Phi}e^{i\th} \right]
 \ ,
\fqr
and is worldsheet diffeomorphism and Weyl invariant.

\section{IIB $(p,q)$ Strings}
\setcounter{equation}{0}

To analyze the theory we choose a (conformally) flat background in 
the Einstein frame $G^E_{\mu\nu}= \eta_{\mu\nu}^E \equiv e^{-\phi/2} 
\eta^{String}_{\mu\nu}$, $B_{\mu\nu}^{(i)}=0$ and a  constant dilaton 
and axion $\phi$ and $\chi$:
\bqr
S^{Pol}_{SL(2,Z)}&=& \int d^2\sig  ~\frac{\rho}{4\pi\alp\pr} 
~\sqrt{g} g^{\alp\bet}\pa_{\alp}X^{\mu}\pa_{\bet}X_{\mu}   
- \rho \cos(\th)e^{\phi/2}\eps^{\alp\bet}F^{(1)}_{\alp\bet}  
\non && 
 - \left[ \rho \cos(\th)
e^{\phi/2}\chi - \rho \sin(\th)e^{-\phi/2} \right] 
 \eps^{\alp\bet} F^{(2)}_{\alp\bet}  \ . 
\label{act22}
\fqr
The preceding steps have linearized the Nambu-Goto form of the action
to the more familiar Polyakov one from which we know in principle how
to compute S-matrix elements of the string fields (e.g. massless fields 
in the gravitational multiplet). 
Note that naively integrating out the world-sheet 
gauge fields forces the Lagrange multiplier field $\rho$ to be constant: 
it yields the tension of the string, just as in the action 
\rf{act1}. We now make this last step more precise and find that
the tension is in fact the $(p,q)$ string tension $T_{pq}$ of equation~ 
\rf{tpq}.  

The above form of the action is similar to that encountered 
in the path integral formulation of T-duality \cite{Buscher:1987,
Buscher:1988, Rocek:1992}.  In the latter one gauges the spacetime
isometry of one of the coordinates, $\del X_T = \eps$, by the
introduction of a worldsheet gaugefield.  One also adds a Lagrange 
multiplier whose field equation enforces the connection to be flat. In
other words the gauge field is pure gauge and there are no new local
degrees of freedom.  After choosing the gauge where the potential vanishes
one recovers the original action; integrating out the gauge field on 
the other hand yields the T-dual action. 

It is this Lagrange multiplier term that we see recurring twice in
the last two terms in \rf{act22}. For each such term\footnote{A boundary
term is required to make this and the following steps exact. This term 
has been suppressed in all of the above. \cite{Townsend:1997, Townsend:1992}} 
\bqr  
\int \ell_i F^{(i)} \qquad \mbox{$i=1,2$}&~~~\mbox{with}~~~& 
F^{(i)}=\epsilon^{\alpha\beta}F^{(i)}_{\alpha\beta}
\label{bacondbble}
\fqr 
we can split the scalar Lagrange multiplier field into a constant zero-mode 
and non-constant piece (see e.g. \cite{Schmidhuber})
\bqr 
\ell_i = \ell^{(0)}_i + \tilde{\ell}_i \qquad\qquad 
\partial_{\alpha}\ell^{(0)}_i =0 ~~;~~
\left.\tilde{\ell}_i\right|_{\sigma,\tau =0} = 0 \ .
\fqr 
On a non-trivial world-sheet single-valuedness of the path-integral
under large gauge transformations forces the 
constant part multiplying the field strength $\ell_i^{(0)}$ to be an 
integer up to an appropriate normalization
\bqr 
\int_{T_2}F^{(i)} ~\epsilon ~Z \qquad\quad  
  \rightarrow \qquad\quad \ell^{(0)}_i ~\epsilon ~Z \ .
\label{whopper}
\fqr 
This quantization follows from the gauge group being $U(1)$ 
rather than ${\bf R}$ \cite{Witten:1996, Schmidhuber}.               

Integration by parts, however, removes the quantized constant part
$\ell^{(0)}$ from the terms (\ref{bacondbble}) 
\bqr 
\int \ell_i F_i = -\int \epsilon^{\alpha\beta} \partial_{\alpha} \ell_i
A^i_{\beta} \equiv -\int \epsilon^{\alpha\beta} \partial_{\alpha}
\tilde{\ell}_i A^i_{\beta} \ .
\fqr 
Now integrating over the gauge field as a Lagrange multiplier forces the
remaining part $\tilde{\ell}_i$ to vanish. This means that the
full coefficient fields $\ell_i$ are constant and quantized; the gauge 
fields have been accounted for as Lagrange multipliers whose constraint 
has been explicitly enforced.

In the context of the action in \rf{act22} the quantized constant
parts are~\cite{Rocek:1992, Witten:1996}   
\bq
\ell_1^{(0)} =\rho_0
\cos(\th_0)e^{\phi/2} = q; \qquad \ell_2^{(0)} = \rho_0 \cos(\th_0)
e^{\phi/2}\chi - \rho_0 \sin(\th_0)e^{-\phi/2} = p \ ,
\fq
or
\bqr
\rho_0  \sin(\th_0) = pe^{\phi/2}+ q\chi e^{\phi/2} ~~&;&~~ 
 \rho_0\cos(\th_0) = qe^{-\phi/2} \ .
\fqr 
In other words the constant part of $\rho$ equals 
\bqr  
\rho_0 &=& e^{\phi/2}\sqrt{(p+ q\chi)^2+q^2 e^{-2\phi} } = 
 2\pi\alpha' e^{\phi/2} T_{pq}\ ,
\fqr
and contains precisely the form of the $SL(2,Z)$ covariantized
tension in eq. 
\rf{tpq}.  Note that as the gauge field $F$ can be extended to the open
string combination $F+B$ the integers $p$ and $q$ are indeed the charges 
under the NS-NS and R-R antisymmetric tensor field.  The extra overall 
factor of $e^{\phi/2}$ is what is required to transform the Einstein metric 
to the string frame. The resulting theory is just that of the 
regular string but with the tension $T$ replaced by its $(p,q)$ quantized 
value $T_{pq}$. 
 
\section{IIB $SL(2,Z)$ Covariant Superstring}
\setcounter{equation}{0}

So far we have discussed an approach to the $SL(2,Z)$
covariant IIB superstring involving the bosonic worldsheet degrees of
freedom only. Our focus, however, is the superstring as it is the 
latter which is expected to possess the $SL(2,Z)$ S-duality. Formally 
the $SL(2,Z)$ covariant action for the IIB superstring is identical 
to the action~\rf{s1} containing only the bosonic fields, except that 
one makes all fields {\em spacetime} supersymmetric 
\cite{Cederwall:1997}. The action is thus again
\bqr
S_{SL(2,Z)}&=& \int d^2\sig~ \frac{\lam^2}{2}\left[ \frac{
 \det(G_{\alp\bet})}{(2\pi\alp\pr)^2} + \Phi\bar{\Phi}\right] \ , 
\fqr
with
\bqr
\Phi &\equiv&\eps^{\alp\bet}
\cU_i\left(F^{(i)}_{\alp\bet}+\frac{B^{(i)}_{\alp\bet}}{2\pi\alp\pr}\right) \ ,
\fqr 
but the pullback of the metric and the antisymmetric tensor
fields are now defined as
\bqr
G_{\alp\bet} &=& \eta_{ab}E_M^aE_N^b\pa_{\alp}Z^M\pa_{\bet}Z^N \ ,\non
B_{\alp\bet}^{(i)} &=& \pa_{\alp}Z^M\pa_{\bet}Z^NB^{(i)}_{MN} \ .
\fqr
Here $Z^M=(X^{\mu},\th_1^{\bf a}, \th_2^{\bf a})$ and $E_M^A=(E_M^a,
E_M^{\bf a_1}, E_M^{\bf a_2})$ are the embeddings of the worldsheet 
into the coordinates and the vielbeins for the target space of 
the IIB superstring.  Supersymmetrizing does not change the 
world-sheet Lagrange multipliers $\rho,\theta$ or the world-sheet 
gauge fields $F^{(i)}$ \cite{Cederwall:1997}.  The 
steps in Section 3 then carry over unaltered to the supersymmetric 
string.  In particular, the Polyakov action 
\bqr
S^{Super}_{SL(2,Z)}&=& \int d^2\sig~ \frac{\rho}{4\pi\alp\pr} \sqrt{g}  
 g^{\alp\bet} \pa_{\alp}Z^M\pa_{\bet}Z^N E_M^aE_{Na}
 - \frac{\rho}{2} e^{-i\th}\Phi - \frac{\rho}{2} e^{i\th}\bar{\Phi}
 \non
&=& \int d^2\sig~ \frac{\rho}{4\pi\alp\pr} \sqrt{g}  
g^{\alp\bet} \pa_{\alp}Z^M\pa_{\bet}Z^N E_M^aE_{Na} 
\non
&& \hspace{-1in}
- \frac{\rho}{2} e^{-i\th}\eps^{\alp\bet}\cU_i\left(F^{(i)}_{\alp\bet}+   
 \pa_{\alp}Z^M\pa_{\bet}Z^N \frac{B^{(i)}_{MN}}{2\pi\alp\pr}\right) 
-\frac{\rho}{2} e^{i\th}\eps^{\alp\bet}\bar{\cU}_i\left(F^{(i)}_{\alp\bet}+ 
 \pa_{\alp}Z^M\pa_{\bet}Z^N \frac{B^{(i)}_{MN}}{2\pi\alp\pr}\right),
\label{sact22} 
\fqr
reduces to that of the Green-Schwarz string with again the tension $T$
replaced by that of the $(p,q)$ string. Indeed, as was noted in
\cite{Cederwall:1997} the terms involving the superfield $B_{MN}$
containing the antisymmetric tensor fields yield the
Wess-Zumino-Witten term in flat superspace \cite{T1,T2}. Note also the
presence of the D-string $\int \chi (F+B)^{(2)}$ coupling \cite{Douglas}.

This action is almost identical to the one proposed by Berkovits for
the manifestly $SL(2,Z)$ covariant IIB string compactified on a
Calabi-Yau three-fold \cite{Berkovits}. The difference is roughly 
that in our case the $SL(2,Z)$ inert Lagrange multiplier $\rho$ also 
multiplies the term involving the supervielbeins.

\section{$SL(2,Z)$ Amplitudes}
\setcounter{equation}{0}

We now turn to the computation of $SL(2,Z)$ invariant scattering
amplitudes.  The 
motivation for working with an $SL(2,Z)$ invariant formulation 
is to obtain $SL(2,Z)$ invariant (and dual) scattering expressions directly.
However, a problem arises in that at present 
we do not yet know how to covariantly quantize the ten-dimensional 
kappa-symmetric GS-string. Following the usual light-cone arguments that 
equate the GS-string with the RNS spinning string we may replace
the tension $T$ with that of the $(p,q)$ string there as well;
alternatively one may quantize the GS-string on the light-cone
directly.   In the RNS form we can then use the standard techniques 
of string perturbation theory. 

A $(p,q)$ string ``vacuum'' arises from solving the field equations
for the $SL(2,Z)$ fields $\rho$ and $F^{(i)}$ and spontaneously breaks the 
$SL(2,Z)$ covariance of the action: each of the possible 
values for $\rho_{0}^{(p,q)}$ is an equally legitimate
solution to the field equations.   In quantizing the 
string action the functional integral over $\rho$ is then equivalent
to  a sum over the contributions of each of the $(p,q)$ strings 
\bqr  
Z^{SL(2,Z)} &=& \int \frac{[dX^{\mu}] [dg]}{\cN} [d\rho] [dA^{(i)}]
~ e^{-S_{SL(2,Z)}}  \non
&=& \sum_{(p,q)'}\int \frac{[dX^{\mu}] [dg]}{\cN}  
~ e^{-T_{pq}S_0} \ ,
\label{part}
\fqr  
where the remaining integration is over the matter and gravitional
superfields on 
the world-sheet; $\cN$ represents the modding out of the local worldsheet
symmetries of the string.   
As the notation indicates we have limited the sum to 
relatively prime pairs.  If $(p,q)$ were not relatively prime then
the tension $T_{pq}$ would be the multiple of that of a $T_{p\pr
q\pr}$ string with coprime charges; as this configuration is a 
superposition of independent $(p,q)$ strings \cite{Schwarz:1995} 
we consider this as overcounting in the functional integration.  In
addition, 
the charges $(p,q)$ should not be simultaneously zero as the action  
vanishes in this case.  

The partition function thus remains invariant under an $SL(2,Z)$ 
transformation though perturbative amplitudes in each individual $(p,q)$ 
sector are not.  $SL(2,Z)$ covariant amplitudes are obtained from the full
partition 
function~\rf{part} after summing over the $(p,q)$ sectors, 
\bqr
A_n^{SL(2,Z)} = \sum_{(p,q)'}\int {[dX_m] [dg]\over \cN_n}
~ e^{-T_{pq}S_0} ~ V_1V_2 \ldots V_n\ .
\label{amp}
\fqr  
For $SL(2,Z)$ invariant vertex operators 
the summation over $\rho$ manifestly leads to $SL(2,Z)$ invariant results 
through perturbation theory in the covariant string: the amplitudes are 
the ($SL(2,Z)$ invariant) sum of $(p,q)$ sub-amplitudes.   As the
$(p,q)$ string only differs in the tension from the fundamental string
the naive choice is to use
the regular perturbative string vertex operators in each sector.
Amplitudes for the $(p,q)$
string are then obtained from those of the
fundamental string by substituting $T=1/2\pi \alp\pr$ with
$T_{pq}$ in accordance with S-duality. 
For graviton scattering  
we use in each $(p,q)'$ sector the standard vertex operator 
for the emission of a gravitational state. 

One might argue that the use of the naive vertex operators is 
incorrect because of the extra fields in the $SL(2,Z)$ covariant 
world-sheet action. For example in~\rf{sact22} the coupling to the 
background Einstein metric includes a factor of the Lagrange multiplier 
field $\rho$.  On-shell this is constant but not $(p,q)$ independent 
and amounts to a normalization condition depending on $(p,q)$. The
vertex operator should  
in principle be compatible with $SL(2,Z)$ invariance and we 
shall take the $(p,q)$ normalization to be unity.  

The four-graviton amplitude found from \rf{amp} is the sum over 
the $(p,q)$ four-graviton amplitudes ,  
\bqr
A_4^{SL(2,Z)} &=& \sum_{(p,q)'} e^{-2\phi} 
 \kap_{pq}^2 E \frac{(\alp\pr_{pq})^3}       
 {\bar{s}\bar{t}\bar{u}}
\frac{\Gam(1-\bar{s})\Gam(1-\bar{t})\Gam(1-\bar{u})}{\Gam(1+\bar{s}) 
\Gam(1+\bar{t})\Gam(1+\bar{u})} 
\label{sumfourpoint}
\fqr  
with the barred invariants given by, 
\bqr 
\bar{s} = \frac{\alp\pr_{pq}}{4}s, ~~&&~~s+t+u=0 \ .
\fqr
Here $\kappa_{pq}$ is the gravitational coupling constant, 
including a $(p,q)$ dependent normalization and $E$ is the 
kinematic factor (e.g. \cite{GreenSchwarz}) depending on the
polarizations (we shall normalize the $p,q$ dependence in $\kappa_{p,q}$ to 
unity).   After transforming to the Einstein frame, where 
$s=\sqrt{\tau_2} s_E$, the $SL(2,Z)$ invariance is seen explicitly.      

Russo \cite{Russo:1998} has analyzed why this $SL(2,Z)$ invariant
amplitude in \rf{sumfourpoint} does not yield complete non-perturbative 
results for string amplitudes.  The form in \rf{sumfourpoint} counts 
the pole of the gravitational exchange (i.e. massless) an
infinite number of times, once from each of the $(p,q)$   
sectors.  Also, when expanded around weak coupling, the expression
contains odd powers of $g_s$ which cannot occur in closed string 
perturbation theory. This is in contrast to the ansatz in
\cite{Russo:1997}, 
\bqr
A_4 &=& e^{-2\phi} \kap^2 E \frac{1}{{s}{t}{u}}
\prod_{(p,q)'}\frac{\Gam(1-\bar{s})\Gam(1-\bar{t})\Gam(1-\bar{u})} 
 {\Gam(1+\bar{s}) \Gam(1+\bar{t})\Gam(1+\bar{u})} \ ,
\label{fourpoint}
\fqr 
which is also given in string frame.
Instead of a summation over the various sectors it
takes the form of a product in such a way that the expression  
contains a single massless pole, corresponding to the exchange of a 
single gravitational mode.\footnote{The 
product is reminiscent of the closed/open string amplitude 
relations at tree-level: $A_4^{closed}=\sum A_4^{open}\bar{A}_4^{open}$.  
However, because the graviton is spin-two it can not arise from an
infinite product of non-zero spin states.} It is unclear how a 
further dressing with the $SL(2,Z)$ peculiar fields could change the 
sum in~\rf{sumfourpoint} into a product. 

The massless excitations of the different $(p,q)$ sectors are not 
independent but should be identified under the duality. The current 
formalism does not contain this prescription, but an analysis of the 
above action similar to that of the D-string in \cite{Kallosh} could 
provide this. The IIB supersymmetry algebra of the D-string action
contains central charges, which means that the massless excitations -
the gravitational multiplet - are projected out \cite{Kallosh}.  
A projector operator, similar to the GSO, may be inserted by hand to 
eliminate the overcounting associated with the massless tree exchange 
(although we do not have a world-sheet construction of this).   
 
After subtraction of the massless poles, both expressions~\rf{sumfourpoint} 
and~\rf{fourpoint} give rise to the same $R^4$ term in the effective 
action. Our result in~\rf{sumfourpoint} bears out the interpretation 
of this term given in \cite{kehagias}.  Aside from the indirect arguments 
of Russo regarding the massless poles and the powers of $g_s$ we cannot 
in this order tell which of the two results yields the correct non-perturbative 
string scattering amplitudes. A comparison could be made at the next 
order in the derivative expansion or for higher point amplitudes. The
formalism presented here allows one to compute these using the
standard rules of string perturbation theory. 
 
It would be interesting to find a prescription for the computation
of the scattering amplitude that leads to the conjecture 
of Russo. For example, instead of summing over the different 
sectors of the solution space of $\rho$, i.e. $\rho^{(p,q)}_0$, we 
could integrate naively with measure factor $\prod d\rho^{(p,q)}_0$
in the spirit of the heuristic proposal in \cite{Russo:1998}.  
A placement of the product within the integration is then crucial.  
Using the form 
\bqr  
\int d\mu_4 \left[ \prod_{(p,q)'} d\rho_0^{(p,q)}\right] 
 e^{-T_{(p,q)} S} ~ \prod_{i=1}^4 V_i(z,{\bar z}) 
\fqr 
gives an infinite tension in the exponential; this occurs 
either including or not including the vertex operators in the product 
(in the latter case one would have to normalize by dividing by 
the infinite helicity product $\prod_{(p,q)'} E/{\bar s}{\bar t} 
{\bar u}$, which does not seem sensible.).   

Though we have not analyzed the analytic structure of~\rf{sumfourpoint} 
in detail, it appears that both
expressions~\rf{sumfourpoint} and \rf{fourpoint} have the fundamental
problem that they do not contain any cuts or poles off the real
axis. At weak string coupling all the excited states of the $(p,q)
\neq (1,0)$ strings are expected to be unstable and the full
non-perturbative four-graviton amplitude should reflect this. To this
extent neither expression can be the full answer but only a first
approximation.  The natural extension of the results here 
would be to consider amplitudes on higher genus surfaces as well; the
amplitude~\rf{sumfourpoint} arose at string tree-level, i.e.
superspherical 
worldsheet.  Similarly, the product form is conjectured to be a tree-level 
result \cite{Russo:1998}.  The higher-genus corrections could generate 
the required cuts.

This poses the fundamental question as to what parameter governs the 
genus expansion in this formulation.  The correct dilaton
coupling in the GS-string is not yet known.  In the RNS formalism the dual 
$(p,q)$ dilaton measures the expansion in each $(p,q)$ sector and 
taking $\phi_{p,q}=\phi$ leads to explicit $SL(2,Z)$ invariant results. The
expression~\rf{sumfourpoint} implictly takes each of the $(p,q)$ 
coupling constants, $e^{-\phi}$, simultaneously 
infinitesimally small.  The amplitude one generates this way is
manifestly S-dual, though the worldsheet genus expansion of the $SL(2,Z)$
string no longer corresponds to a perturbative expansion in $g_s$. It is 
not clear whether there exists an extension of the dilaton that is able 
to govern the world sheet expansion of the $SL(2,Z)$ covariant string
in an invariant way. This problem has been solved by the introduction 
of additional fields for the manifestly four dimensional formulation 
of the IIB $SL(2,Z)$ covariant superstring by Berkovits \cite{Berkovits} 
but a corresponding formalism is as yet unknown for the ten-dimensional case.

\section{Conclusion}
\setcounter{equation}{0}

In summary we have shown that the $SL(2,Z)$ covariant action for the IIB
string
of \cite{Townsend:1997, Cederwall:1997} may be rewritten in a Polyakov
form. This is a first step on the way to quantization and manifestly 
organizes the perturbative expansion in an $SL(2,Z)$ invariant fashion.  
An analysis of the worldsheet field equations shows that the
groundstates of the worldsheet theory correspond to individual $(p,q)$
strings.  One is required to sum over the various sectors in the first 
quantized partition function, with each term corresponding to the 
contribution for the individual $(p,q)$ superstring. $SL(2,Z)$ covariant 
four- and higher point amplitudes can be computed following this approach 
by taking the sum over the individual $(p,q)$ ones.  We find the 
four-graviton scattering amplitude; it reproduces
the modular $f(\tau,\bar\tau)$ coefficient of the $R^4$ term in 
the low energy effective action of \cite{Green:1997one}.  The
conjectured $SL(2,Z)$ covariant amplitude of Russo does so as well but
the latter contains a single massless pole, which is as one expects,
whereas ours counts an infinite number before projection, one 
from each sector. These 
massless poles ought to be identified under the action of $SL(2,Z)$.  A 
problem is the identification of the appropriate small parameter that 
measures the world-sheet genus expansion in an $SL(2,Z)$ invariant way. 
This should be an extension of the $(p,q)$ dilaton which governs the 
expansion for each $(p,q)$ sector.

\vskip .2in
{\bf Acknowledgements}

\noi
We would like to thank M. Green, S. Sethi and M. Ro\v{c}ek for 
discussions.  K.S. would especially like to thank Paul 
Townsend for discussions during the String theory workshop in
Amsterdam, whose organizers he wishes to thank. K.S.
is also grateful to the organizers of the NATO ASI ``Quantum
Geometry'' in Akureyri for their hospitality and an opportunity to
present results of this work.  G.C. thanks the Aspen Center for Physics 
for hospitality during which time this work was completed.  This 
work is supported in part by the U.S. Department of Energy, Division 
of High Energy Physics, Contract W-31-109-ENG-38.

\end{document}